\documentstyle[prl,aps,twocolumn,mycite,mypsfig2]{revtex}

\def\>{\rangle}

\def\ua{$\uparrow$}
\def\da{$\downarrow$}

\begin{document}

\title{Liquid state NMR Quantum Computing}
\date{\today}
\author{Lieven M.K. Vandersypen, Costantino S. Yannoni, Isaac L. Chuang}
\maketitle

\def\be{\begin{equation}}
\def\ee{\end{equation}}
\newcommand{\ket}[1]{\mbox{$|#1\rangle$}}
\newcommand{\mypsfig}[2]{\psfig{file=#1,#2}}

\tableofcontents

\clearpage
\section{Introduction}

Since its invention, NMR spectroscopy has developed from a technique
for studying physical phenomena such as magnetism into a tool for
acquiring information about molecules in chemistry and
biology. Furthermore, it was pointed out early on (1955), almost as an
anecdote, that nuclear spins could also be used for storing
information using spin echoes~\cite{anderson55}.  

This insight
beautifully illustrated a notion that was developed in a very
different context: information is physical and cannot exist without a
physical representation~\cite{landauer}.  In recent decades, the
relationship between physics and information has been revisited from a
new perspective: could the laws of physics play a role in {\em how}
information is processed?  The answer appears to be yes. If
information is represented by systems governed by the laws of quantum
mechanics, such as nuclear spins, an entirely new way of doing
computation, quantum computation (QC), becomes possible. Quantum
computing is not just different or new; it offers an extraordinary
promise, the capability of solving certain problems which are beyond
the reach of any machine relying on the classical laws of physics!

The practical realization of quantum computers is
still in its infancy. Interestingly, over 40 years after the initial
suggestion of using spins to represent (classical) information, NMR
spectroscopy has become the workhorse for experimental explorations of
quantum information processing.

In this article, we first explain how quantum computers work and why
they could solve certain problems so much faster than any classical
computer.  Next, we describe how quantum computers can be implemented
using NMR techniques and what is involved in designing and
implementing QC pulse sequences, preparing a suitable initial state
and interpreting the output spectra. We close with an overview of the
state-of-the-art and the prospects for NMRQC and other QC
implementations.

Good reviews of quantum computation and information can be found in Refs.
~\incite{bennett00,ekert96,chuang00}. A good intro to and a specialized review of NMRQC are given in Ref. ~\incite{jones00a} and Ref. ~\incite{gershenfeld98} respectively.

\section{Quantum computation}
\label{sec:qc}

\subsection{INEPT as a computation}
\label{sec:inept}

We start from a familiar place for many NMR spectroscopists, the INEPT
pulse sequence ({\em see \bf INEPT}). This sequence was designed to
transfer polarization from a high $\gamma$ nucleus to a low $\gamma$
nucleus. However, it can also be viewed as a {\em logic gate}
(Fig.~\ref{fig:inept}) which flips one spin conditioned upon the
orientation of a neighbouring spin.

If we arbitrarily assign ``0'' to a spin up and ``1'' to a spin down,
we can think of spin-1/2 nuclei as bits in a digital computer. We
remind the reader that bits (``0'' or ``1'') can be used to represent
numbers in a similar way to decimal numbers (``0'' through ``9''). In
decimal representation, $503$ means $3 \times 10^0 + 0 \times 10^1 + 5
\times 10^2$; similarly, in binary representation, $1101$ means $1
\times 2^0 + 0 \times 2^1 + 1 \times 2^2 + 1 \times 2^3$ (which
corresponds to the number $13$ in decimal representation). All
computers today process information in binary representation, and are
able to rapidly perform multiplication and division (repeated addition
and substraction) as well as the most complex computations, via a
sequence of very simple, elementary operations, called logic gates,
acting on just one or two bits at a time.

In this binary framework, the INEPT sequence corresponds to an
elementary two-bit operation known as the controlled-{\sc not} or for
short {\sc cnot} gate (within the phase corrections discussed in
Section~\ref{sec:design}). The {\sc cnot} gate performs a {\sc not}
operation on one bit, flipping it from ``0'' to ``1'' or from ``1'' to
``0'', if and only if the value of a second bit is ``1''.  The input
to the logic gate is the initial state of the spins, and the output is
the final state of the spins. The four possible input values and the
corresponding output values are tabulated in Fig.~\ref{tab:inept}.

The {\sc cnot} combined with single-spin rotations provides for a {\em
universal} set of logic gates. This means that {\em any} computational
task can be implemented using a sufficiently large number of nuclear
spins simply by concatenating {\sc cnot}s and single-spin rotations in the
proper way~\cite{gershenfeld97,cory97,cory98a}. In summary,\\

\indent {\em spin-1/2 nuclei in a molecule can serve as bits in a
computer, and pulses and delay times provide universal logic gates.}\\

\subsection{Quantum parallellism}
\label{sec:qubits}

Everything we discussed so far was purely classical. The Bloch-sphere
picture of Fig.~\ref{fig:inept} reinforces this classical view of the
spins; however, nuclear spins are really quantum objects. In Dirac
notation, the state of a spin can be denoted $|0\>$ for a spin in the
ground state (along $z$), and $|1\>$ for a spin in the excited
state (along $-z$), corresponding to the two classical values
for a bit (``0'' and ``1''). Now, a spin said to be 'along the
$x$ axis' is in reality in a coherent superposition state of
spin up and spin down, written as $(|0\> + |1\>)/\sqrt{2}$, a spin
'along the $y$ axis' is in the state $(|0\> + i |1\>)/\sqrt{2}$,
etcetera. A spin-1/2 particle is thus more than just an ordinary
bit.\\

\indent {\em Any two-level quantum system, such as a spin-1/2
particle, can serve as a quantum bit (qubit).}\\

The difference between the quantum and classical descriptions becomes
clear as soon as more than one quantum particle is
considered. For example, it is well-known that the state of $n$
interacting spins-1/2 cannot be described simply by $n$ sets of
coordinates on the Bloch sphere.  In order to include phenomena such
as multiple-quantum coherence, we need recourse to $4^n-1$ real
numbers in the product operator expansion or equivalently to density
matrices of dimension $2^n \times 2^n$. Furthermore, the evolution of
a closed system of $n$ spins can only be described by $2^n \times 2^n$
unitary matrices ({\em see \bf Liouville Equation of Motion}).  The
number of degrees of freedom that need to be specified in a classical
description of the state and dynamics of $n$ coupled spins thus
increases exponentially with the number of spins.

Richard Feynman proposed in 1982 that the exponential complexity of
quantum systems might be put to good use to simulate the
dynamics of another quantum system~\cite{feynman82}, a task which
requires exponential effort on a classical computer. David Deutsch
extended and formalized this idea in 1985, and introduced the notion
of ``quantum parallellism''~\cite{deutsch85}.

Consider a (classical) logic gate which implements a function $f$ with
one input bit $x$ and one output bit $f(x)$. If $x=0$, the gate will
output $f(0)$; if $x=1$, the output will be $f(1)$.  The analogous
quantum logic gate is described by a unitary operation which
transforms a qubit as
$$
|0\> \mapsto |f(0)\> \;\;\; \mbox{and} \;\;\; |1\> \mapsto |f(1)\> \,.
$$
However, due to the possibility of preparing coherent superposition
states and to the linearity of quantum mechanics, the same gate also
performs the transformation
$$
\frac{|0\> + |1\>}{\sqrt{2}} \;\; \mapsto \;\; \frac{|f(0)\> +
|f(1)\>}{\sqrt{2}} \,.
$$
In this sense, it is possible to evaluate $f(x)$ for both input values
in one step!  Next consider a different logic gate which implements a
function $g(x)$ with two input bits. We can prepare each qubit in a
superposition of ``0'' and ``1''. Formally, the state of the joint
system is then written as $(|0\> + |1\>) \otimes (|0\> + |1\>)/2$.
Leaving out the tensor product symbol as well as any normalization
factors, the state can be written as $(|0\> + |1\>)(|0\> + |1\>)$, or
$|0\>|0\>+|0\>|1\>+|1\>|0\>+|1\>|1\>$, which is further abbreviated to
$|00\> + |01\> + |10\> + |11\>$. Therefore, a set of two spins can be
in a superposition of the four states ``00'', ``01'', ``10'' and
``11''. A quantum logic gate implementing $g(x)$ then transforms this
state as
$$
|00\> + |01\> + |10\> + |11\> \mapsto |g(00)\> + |g(01)\> + |g(10)\> +
 |g(11)\>
$$
Thus the function has been evaluated for the four possible input
values in parallel.  In general, a function of $n$ qubits implemented
on a quantum computer can be evaluated for all $2^n$ input values in
parallel! In contrast to classical computers, for which the number of parallel
function evaluations increases at best linearly with their size, \\

\indent {\em the number of parallel function evaluations grows
exponentially with the size of the quantum computer (the number of
qubits).} \\

Obviously, this is true only so long as the coherent superposition
states are preserved throughout the computation. This means that the
computation should be completed before quantum coherence is lost due
to ``decoherence'' processes (in NMR spin-spin and spin-lattice
relaxation; {\em see \bf Relaxation: An Introduction}). Since
some degree of decoherence is unavoidable, practical quantum computers
appeared virtually impossible to build, until quantum error
correction was conceived, as discussed in Section~\ref{sec:qec}.

Even if the coherence time is long compared to the duration of a
typical logic gate and quantum error correction is employed, can we
really access the exponential power exhibited by quantum systems? The
postulates of quantum mechanics dictate that a measurement --- which
induces instantaneous and complete decoherence --- of a qubit in a
superposition state $|f(0)\> + |f(1)\>$ will give either ``f(0)'' or
``f(1)'', with equal probabilities.  Similarly, after doing $2^n$
computations all at once, resulting in a superposition of $2^n$
output values, measurement of the quantum bits randomly returns a
single output value. A more clever approach is thus needed: \\

\indent {\em exploiting quantum parallellism requires the use of
quantum algorithms.}

\subsection{Quantum algorithms}
\label{sec:algorithms}

Special quantum algorithms allow one to take advantage of quantum
parallellism in order to solve certain problems in far fewer steps
than is possible classically. When comparing the capability of two
computers to solve a certain type of problem,\\

{\em the relevant criterion is not so much what resources (time, size,
signal-to-noise ratio, $\ldots$) are required to solve a specific
instance of the problem but rather how quickly the required resources
grow with the problem size.}\\

A particulary important criterion is whether the required resources
increase exponentially or polynomially with the problem
size. Exponentially difficult problems are considered intractable ---
they become simply impossible to solve when the problem size is
large. In contrast, polynomially difficult problems are considered
tractable or possible to solve. The interest in quantum computing is
based on the fact that\\

{\em certain problems which appear intractable (resources grow
exponentially with problem size) on any classical computer are
tractable on a quantum computer.}\\

This was shown in 1994 by Peter Shor, almost 10 years after Deutsch
introduced quantum parallellism. Shor's quantum
algorithm~\cite{shor94} allows one to find the period of a function
exponentially faster than any classical algorithm. The importance of
period-finding lies in that it can be translated, using some results
from number theory, to finding the prime factors of integer numbers,
and thus also to breaking widely used cryptographic codes. These codes
are based precisely on the fact that no {\em efficient} classical
algorithm is known for period-finding or factoring, i.e. the effort
required to factor a number on classical computers increases
exponentially with the number of digits of the integer to be factored.
In contrast, Shor's algorithm is efficient: on a quantum computer, the
effort to factor an integer increases only polynomially (to be
precise, the third power) with the number of digits of the integer. As
a result, while factoring a 1000-digit number is believed to be beyond
the reach of any machine relying on the classical laws of physics,
such a feat could be accomplished on a quantum computer.

The first quantum algorithm was invented by Deutsch and
Jozsa~\cite{deutsch92} (1992).  This algorithm allows a quantum
computer to solve with certainty an artificial mathematical problem
known as Deutsch's problem. Even though this algorithm does not have
much application, it is historically significant as it provided the
first steps towards Shor's algorithm, and because it is a simple
quantum algorithm that can be experimentally tested.

Another class of quantum algorithms was discovered in 1996 by Lov
Grover.  These algorithms~\cite{grover96} allow quadratic speed-ups of
certain search problems, for which there is no better approach
classically than to try all $L$ candidate solutions one at a time. A
quantum computer using Grover's algorithm needs to make only
$\sqrt{L}$ such trials.  Even though this speed-up is only quadratic
rather than exponential, it is still significant.

The last currently known application of quantum computers lies in
simulating other quantum systems~\cite{lloyd96}, as Feynman
conjectured. Even a computer consisting of no more than a few dozen
quantum bits could outperform the fastest classical computers in
solving important physics problems, such as calculating the energy
levels of an atom~\cite{abrams99}.\\

In the remainder of this section, we will briefly review the structure
of Shor's algorithm, because it is so important and at the same time
gives good insight into how quantum computing works (for a more
detailed explanation, see Refs. ~\incite{ekert96} and
~\incite{shor94}).  The crucial step in Shor's factoring algorithm is
the use of the quantum Fourier transform (QFT) to find the period $r$
of the function $f(x) = a^x \mbox{mod} M$, which means $f(x)$ is the
remainder after division of $a^x$ by $M$, where $M$ is the integer to
be factored, and $a$ is an integer which is more or less randomly
chosen~\cite{ekert96,shor94}.

The QFT performs the same transformation as the (classical) fast
Fourier transform (FFT), but can be computed exponentially faster. As
always, we don't have access to all the individual output values; the
QFT merely allows us to {\em sample} the FFT but as we will see this
suffices for period-finding. The FFT$_N$ takes as input a string of
$N$ complex numbers $x_j$ and produces as output another string of $N$
complex numbers $y_k$, such that
\be y_k = \frac{1}{\sqrt{N}} \sum_{j=0}^{N-1} x_j e^{2\pi i jk/N} \,.
\ee 
For an input string with numbers which repeat themselves with
period $r$, the FFT$_N$ produces an output string with period $N/r$,
as illustrated in the following four examples for $N=8$ (the output
strings have been rescaled for clarity)
$$
\begin{tabular}{ccccccccccccccccccccccccc}
$r\;\;$ & &\multicolumn{8}{c}{input string} &&&& \multicolumn{8}{c}{output string}& & $\;N/r$ \,\,& \\ 
8$\;\;$ & &1&0&0&0&0&0&0&0& &$\mapsto$ & &1&1&1&1&1&1&1&1& & 1 & \,\,(a) & \\ 
4$\;\;$ & &1&0&0&0&1&0&0&0& &$\mapsto$ & &1&0&1&0&1&0&1&0& & 2 & \,\,(b) & \\
2$\;\;$ & &1&0&1&0&1&0&1&0& &$\mapsto$ & &1&0&0&0&1&0&0&0& & 4 & \,\,(c) & \\
1$\;\;$ & &1&1&1&1&1&1&1&1& &$\mapsto$ & &1&0&0&0&0&0&0&0& & 8 & \,\,(d) & \\
\end{tabular}
$$
If $r$ does not divide $N$, the inversion of the period is approximate.
Furthermore, the FFT turns shifts in the locations of the numbers in the
input string into phase factors in front of the numbers in the output
string:
$$
\begin{tabular}{cccccccccccccccccccccc}
& 1& 0& 0& 0& 1& 0& 0& 0& &$\mapsto$ & & 1& 0& 1& 0& 1& 0& 1& 0& \,\,(e) & \\
& 0& 1& 0& 0& 0& 1& 0& 0& &$\mapsto$ & & 1& 0&-i& 0&-1& 0& i& 0& \,\,(f) & \\
& 0& 0& 1& 0& 0& 0& 1& 0& &$\mapsto$ & & 1& 0&-1& 0& 1& 0&-1& 0& \,\,(g) & \\
& 0& 0& 0& 1& 0& 0& 0& 1& &$\mapsto$ & & 1& 0& i& 0&-1& 0&-i& 0& \,\,(h) & \\
\end{tabular}
$$

The QFT performs exactly the same transformation, but differs from the
FFT in that the complex numbers are stored in the amplitude and phase
of the terms in a superposition state. The amplitude of the $|000\>$
term represents the first complex number, the amplitude of the
$|001\>$ term the second number and so forth. For clarity, we will
label the states $|000\>, |001\>,\ldots |111\>$ as $|0\>, |1\>, \ldots
|7\>$. As an example, we see from (f) that the QFT transforms the
state $|1\> + |5\>$ to the state $|0\> - i |2\> - |4\> + i |6\>$.

The QFT is incorporated in actual quantum algorithms as outlined in
Fig.~\ref{fig:shor_structure}. A first register (group of qubits) is
prepared in a superposition of all its possible states. A second
register is initialized to the ground state (for factoring a number
$M$, the size of the second register must be at least $\log_2 M$ and
the first register must be at least twice as large). For example, if
register 1 has three qubits and register 2 has two qubits, the state
of the system is prepared in \be
(|0\>+|1\>+|2\>+|3\>+|4\>+|5\>+|6\>+|7\>) |0\> \,.
\label{eq:input_state}
\ee

Then the function $f(x)$ is evaluated (in NMR by applying a pulse
sequence, as discussed in Sections~\ref{sec:inept}
and~\ref{sec:design}), where $x$ is the value of the first register,
and the output value $f(x)$ is stored in the second register. {\em
Since the first register is in an equal superposition of all $|x\>$,
the function is evaluated for all values of $x$ from $0$ to $7$ in
parallel}. For example, let $f(x) = 3$ for even $x$ and $f(x) = 1$ for
odd $x$, which means the period $r$ is $2$ (in real applications, we
have a description of $f$ but do not know $r$ in advance). Evaluation
of $f(x)$ then transforms the state of eq.~\ref{eq:input_state} to the
state
$$
\hspace*{-0.5in} |0\>|3\>+|1\>|1\>+|2\>|3\>+|3\>|1\>
$$
\be
\hspace*{0.5in} +|4\>|3\>+|5\>|1\>+|6\>|3\>+|7\>|1\> 
\ee 
\be =
(|0\>+|2\>+|4\>+|6\>)|3\> + (|1\>+|3\>+|5\>+|7\>)|1\> \,.  
\label{eq:entangled}
\ee 
We pause to point out that this state is {\em entangled}, which means
that it cannot be written as a product of single-qubit states.  The
state $\ket{00}+\ket{01}+\ket{10}+\ket{11}$ is an example of an
unentangled state, because it can be written as
$(\ket{0}+\ket{1})(\ket{0}+\ket{1})$, a product of single-qubit
states. In contrast, $\ket{00} + \ket{11}$ is a simple example of an
entangled state. Entanglement has no classical analogue and is
believed to lie at the heart of the exponential speed-up offered by
quantum computation.

In order to appreciate the role of the QFT, suppose we now measure the
second register in Eq.~\ref{eq:entangled} (this measurement can be
left out but simplifies the explanation). The state of the first
register will collapse to either
\be |0\>+|2\>+|4\>+|6\>
\hspace*{2ex} \mbox{or} \hspace*{2ex} |1\>+|3\>+|5\>+|7\> \,, 
\label{eq:post_meas}
\ee
depending on whether the measurement of register $2$ gave ``3'' or
``1''. We see that all eight possible outcomes are still equally
likely, so the result of a measurement does not give us any useful
information. But if we apply the QFT, the first register will be
transformed to \be |0\> + |4\>
\hspace*{3ex} \mbox{or} \hspace*{3ex} |0\> - |4\> \,.
\label{eq:output_qft}
\ee
Now a measurement of the first register does give useful information,
because only multiples of $N/r$ are possible outcomes, in this example
``0'' and ``4''. 

This concludes the quantum part of the computation. From the
measurement result, a classical computer can efficiently calculate the
inverted period $N/r$, and thus also $r$, with high probability of
success using results from number theory. Now that $r$ is known,
the factors of the integer $M$ can be quickly computed as well, with
high probability (the probability of success can be further increased
by repeating the whole procedure a few times).

We close with two final remarks on quantum algorithms. (1) Quantum
computing cannot offer any speed-up for many common tasks, such as
adding up two numbers or word processing, which can already be
completed efficiently on a classical computer. (2) There are many
exponentially difficult problems which no currently available quantum
algorithm could help solve faster than is possible classically. It
would be somewhat disappointing from a practical viewpoint if no other
applications were found; however, our understanding of the connection
between physics and information and computation has already changed
dramatically.

\subsection{Quantum error correction}
\label{sec:qec}

Any quantum computation must be completed within the coherence time,
in NMR $T_2$ and $T_1$, as pointed out in Section~\ref{sec:qubits}.
$T_1$ and $T_2$ processes alter the state of the qubits and are
therefore a source of errors. For many years, this requirement led to
wide-spread pessimism about the practicality of quantum computers.
In 1995, however, Peter Shor and Andrew Steane independently discovered
quantum error correction~\cite{shor95,steane96} and showed that \\

{\em It is possible to correct for truly random errors caused 
by decoherence.}\\

This came as quite a surprise, because quantum error correction had to
overcome three important obstacles: (1) the no-cloning theorem, which
states that it is not possible to copy unknown quantum
states~\cite{chuang00}, (2) measuring a quantum system affects its
state and (3) errors on qubits can be arbitrary rotations in Hilbert
space, compared with simple bit flips in classical computers.  Quantum
error correction requires many extra operations and extra qubits
(ancillae), though, which might introduce more errors than
are corrected, especially because the effect of decoherence increases
exponentially with the number of entangled qubits, much in the same
manner that multiple quantum coherences decay exponentially faster
than single quantum coherences. Therefore, a second surprising
result~\cite{aharonov99} was that \\

{\em provided the error rate (probability of error per elementary
operation) is below a certain threshold, and given a fresh supply of
ancilla qubits in the ground state, it is possible to perform
arbitrarily long quantum computations.}\\

The threshold error rate is currently estimated ~\cite{aharonov99} to
be about $0.001 \%$. The actual error rate in NMRQC is approximated by
$1 / 2 J T_2$, where $2 J T_2$ is roughly the number of operations
that can be computed within the coherence time. For small molecules
the error rate is typically on the order of $0.1 \%$ to $1 \%$,
two to three orders of magnitude too high. But a remarkable
implication of quantum error correction is that if (1) a molecule is
found which achieves the accuracy threshold and (2) select spins can
be fully polarized, both $T_1$ and $T_2$ could in principle be
infinitely lenghtened by applying an error correction pulse sequence.

\section{NMR quantum computers}

\subsection{Pulse sequence design}
\label{sec:design}

The translation of abstract quantum algorithms or function evaluations
into actual pulse sequences may appear obscure at first sight.
However, systematic techniques~\cite{barenco95,price99} exist to make
pulse sequence design relatively straightforward.  The starting point
is that\\

\indent {\em each quantum algorithm can be described by a sequence of
transformations under unitary operators.}\\

Such unitary transformations represent rotations in Hilbert space (a
multidimensional extension of the Bloch-sphere). Examples of unitary
transformations are evolution during RF pulses and free evolution
under the system Hamiltonian; relaxation processes give rise to
non-unitary transformations. Once the desired unitary operators have
been identified,\\

{\em arbitrary unitary operators can be translated into sequences of
single-qubit rotations and {\sc cnot} gates.}\\

These building blocks can be readily implemented in NMR (see
Section~\ref{sec:implement}). Decompositions into other sets of
elementary gates are also possible, and can be helpful in simplifying
the pulse sequences~\cite{jones98c,vandersypen99}. In any case, it is
crucial that the duration of the pulse sequence design process as well
as the length of the resulting pulse sequence not increase
exponentially with the problem size.

We now point out two important distinctions between QC and
conventional pulse sequences. On the one hand, QC sequences must be
more general:\\

{\em QC sequences must perform the desired transformation for
arbitrary input states.}\\

In contrast, conventional sequences are often designed assuming a
particular input state. As a first example of this difference, the
sequence of Fig.~\ref{fig:inept} assumes that both spins are in Zeeman
states, i.e. aligned along $\pm z$. It implements the unitary operator

\be
\hat{U}_{\mbox{\small INEPT}}= \left(\matrix{1 & 0 & 0 & 0 \cr
				0 & i & 0 & 0 \cr
				0 & 0 & 0 & 1 \cr
				0 & 0 &-i & 0 }\right) \,,
\ee

\noindent which is similar to but different from the unitary operator for the
{\sc cnot} gate, defined as

\be
\hat{U}_{\mbox{\small {\sc cnot}}} = \left(\matrix{1 & 0 & 0 & 0 \cr
				0 & 1 & 0 & 0 \cr
				0 & 0 & 0 & 1 \cr
				0 & 0 & 1 & 0 }\right) \,,
\ee

\noindent implemented, for example, by $90_z^a \, 90_{-z}^b \, 90_x^b \,
1/2J_{ab} \, 90_{-y}^b$.

As a second example, consider the so-called Hadamard gate, defined as
\be
\hat{U}_{\mbox{\small Had}} = \frac{1}{\sqrt{2}}
	\left(\matrix{1 & 1 \cr
		      1 &-1 }\right) \,.
\label{eq:Had}
\ee 

This gate creates a superposition state starting from a basis state:
it transforms $|0\>$ to $|0\>+|1\>$ ($z$ to $x$ in the Bloch sphere)
and $|1\>$ to $|0\>-|1\>$ ($-z$ to $-x$). At first sight, this
transformation could be done simply via a $90_y$ pulse. However, the
unitary operator for $90_y$

\be \hat{U}_{90_y} = \frac{1}{\sqrt{2}} \left(\matrix{1 &-1 \cr 1 & 1
}\right) \,,
\label{eq:90y}
\ee 

\noindent is different from $\hat{U}_{\mbox{\small Had}}$; e.g. applying
$\hat{U}_{\mbox{Had}}$ twice has no net effect, but applying
$\hat{U}_{90_y}$ twice produces $U_{180_y}$. A possible sequence to
implement $\hat{U}_{\mbox{\small Had}}$ exactly is $90_y 180_x$.\\

On the other hand, QC sequences can be more specific:\\

{\em QC sequences can be specialized for a specific molecule using
full knowledge of its spectral properties.}\\

In contrast, conventional sequences must work for any molecule,
because the spectral properties of the molecule are usually not known
in advance. Exact knowledge of the chemical shifts and $J$-coupling
constants allows one not only to greatly simplify the pulse sequences,
but also to achieve much more accurate unitary transformations than
would otherwise be possible.

Finally, while systematic procedures exist to design a pulse sequence,
there is a need to develop tools to find the pulse sequence of the
shortest duration and with the fewest RF pulses. Even small-scale
quantum computations easily involve tens to hundreds of gates acting
on multiple spins and precise control of the spin dynamics is
difficult to maintain throughout such long sequences of operations, as
shown in the next section.

\subsection{Implementation of computations}
\label{sec:implement}

The implementation of quantum computations with NMR can be based on
single-spin rotations and {\sc cnot} gates, since any quantum algorithm can
be translated into these building blocks. Although these elementary
operations appear quite easy to implement, \\

{\em the requirements for precision in QC experiments are unusually high,
due to the large number of pulses and the quantitive nature of
the information contained in the output spectra.}\\
 
Implementation of accurate {\em single-spin rotations} about an axis
in the $x$-$y$ plane is relatively easy in heteronuclear
molecules; yet, it can be very demanding for homonuclear spin systems
because spin selectivity requires longer pulses resulting in
considerable coupled evolution of the spins during the
pulses~\cite{linden99b}.  Clearly, some degree of homonuclearity is
unavoidable when more than a handful of qubits are involved.

We therefore begin by reviewing the requirements for pulse
shaping~\cite{freeman97} ({\em see \bf Shaped Pulses; Selective
Pulses}).  First, the magnetization corresponding to each of the lines
in a multiplet must be rotated about exactly the same axis and over
exactly the same angle, i.e. off-resonance effects due to line
splitting must be removed. This requires self-refocusing shaped pulses
or tailored composite pulses~\cite{cummins99} ({\em see \bf Composite
Pulses}).  Second, the effect of $J$-couplings between unselected
spins must be removed, either during the pulse or later in the pulse
sequence.  Third, all pulses must be universal rotors, i.e. the
rotation must be independent of the initial state of the spin.
Fourth, the unselected spins must not be affected by the RF
irradiation. This last requirement is difficult to satisfy because of
transient Bloch-Siegert effects~\cite{emsley90} which can result in
substantial (tens of degrees) phase shifts of nearby unselected
spins. However, it is possible to estimate and compensate for the
Bloch-Siegert shift~\cite{knill00,vandersypen00b}.  Finally,
simultaneous (as opposed to consecutive) pulses at two or more nearby
frequencies are desirable in order to keep pulse sequences short, but
transient Bloch-Siegert shifts greatly deteriorate such simultaneous
rotations~\cite{kupce95,linden99a}. Nevertheless, accurate
simultaneous rotations at nearby frequencies can be achieved using a
special correction technique~\cite{steffen00}. Still, simultaneous
pulses on well-coupled spins may excite multiple quantum
coherences~\cite{kupce94}.

There are a number of hardware requirements for successful execution
of QC experiments. Good RF coil homogeneity is crucial in avoiding
excessive signal attenuation and related errors. Furthermore, it is
desirable that one frequency source and transmitter board be available
per qubit. If there are more qubits than spectrometer channels, the
carrier frequency must be jumped to the appropriate frequencies
throughout the pulse sequence, or phase ramping techniques must be
employed~\cite{patt92}. A dedicated frequency source for each qubit
also makes it easy to keep track of the rotating frame of each spin
and apply all the pulses on any given spin with the correct relative
phase.  This removes the need to refocus chemical shift
evolution~\cite{freeman97}, which involves extra
pulses. Alternatively, software rotating frames can be created by
detailed bookkeeping of the time elapsed since the beginning of the
pulse sequence such that the evolution of the rotating frame of any
given spin with respect to the carrier reference frame can be
calculated. The phases of the pulses throughout the pulse sequence, as
well as the receiver phase, can then be adjusted
accordingly~\cite{vandersypen00a,knill00}.  The same technique can be
used to implicitly realize single-spin rotations about
$z$. Alternatively, $z$-rotations can be implemented using resonance
offsets or composite pulses~\cite{freeman97}.

Two strategies exist for implementing {\sc cnot} gates (both assume
first order spectra). If all the spins are mutually coupled, {\sc
cnot}'s can be realized via line-selective pulses which invert
specific lines within a multiplet~\cite{cory98a}. In practice, it is
usually more convenient to use pulse sequences such as the one in
Fig.~\ref{fig:inept}~\cite{gershenfeld97,cory98a}.  For molecules with
several coupled spins, the sequence of Fig.~\ref{fig:inept} must be
expanded with extra pulses to refocus the undesired $J$-couplings;
systematic methods exist to design good refocusing
schemes~\cite{leung00,jones99a,linden99a}.  A {\sc cnot} between two
uncoupled spins can be realized by swapping qubit
states~\cite{lloyd93,collins00}. For example, for a {\sc cnot} between
two spins $a$ and $c$ which are not mutually coupled but which are
both coupled to a third spin $b$, the procedure is as follows: apply a
pulse sequence which swaps the state of $a$ and $b$ (via {\sc
cnot}$_{ab}$ {\sc cnot}$_{ba}$ {\sc cnot}$_{ab}$), then perform a {\sc
cnot} between $b$ and $c$, and then swap $a$ and $b$ again. The net
result is {\sc cnot}$_{ac}$; spin $b$ is unaffected. It is thus not
necessary that all spins be pairwise coupled as long as the network
of couplings includes all $n$ spins.

An alternative to imposing the correct evolution on all the spins at
all times, both for RF pulses and for {\sc cnot}-type gates, is to allow
erroneous evolutions which will later be reversed. Such techniques
have been highly successful in certain standard pulse sequences ({\em
see \bf Decoupling Methods}), but are much harder to develop for the
non-intuitive and non-transparent QC pulse sequences. Nevertheless, it
has been shown experimentally that a large degree of cancellation of
erroneous evolutions is possible even in QC experiments: about 300
two-qubit gates involving over 1350 RF pulses have been successfully
concatenated~\cite{vandersypen00a}. A general methodology designed to
take advantage of this possibility has yet to be developed.

From this discussion, it will be clear that\\ 

{\em the selection of a suitable molecule is crucial for
\mbox{NMRQC}}. \\

The desired properties are (1) sufficiently large chemical shifts for good
addressability, (2) large coupling constants, while maintaining
first-order spectra, for fast two-qubit gates (or a coupling network
that matches the pattern of connectivities needed for the algorithm),
and (3) long $T_2$'s and $T_1$'s in order to allow time to execute
many logic gates. Furthermore, high-$\gamma$ nuclei are desirable for
good sensitivity. More mundane but equally important requirements are
that the molecule be stable, synthesizable, soluble and safe.

\subsection{State initialization}

Apart from experiments designed to produce non-thermal spin
polarizations, setting up a proper initial state for the nuclear spins
is a concept worth revisiting for NMR spectroscopists. Since this is a
crucial step in quantum computing, this entire section is therefore
devoted to state initialization.

Most quantum computations require a {\em pure} initial state, for example a
set of fully polarized spins, in the state $\ket{00\ldots0}$.  However,
nuclear spins in thermal equilibrium at room temperature are in an almost
fully random state: for typical magnetic field strengths, the ground
($|0\>$) and excited ($|1\>$) state probabilities differ by only about $1$
part in $10^5$. The spins are then said to be in a {\em mixed} (non-pure)
state.  The polarization could be increased using hyperpolarization
techniques ({\em see \bf Unusual signal enhancement: optical pumping $\&$
hyperpolarized inert gases \em and \bf Optically enhanced magnetic
resonance}) but the state of the art is still very far from cooling nuclear
spins into the ground state.

The conceptional breakthrough which made NMR quantum computation
possible at room temperature was the the concept of {\em
effective pure} or {\em pseudo-pure}
states~\cite{gershenfeld97,cory97,cory98a}:\\

{\em effective pure states are mixed states which produce the same
signal as a pure state to within a scaling factor.}\\

The signature of an effective pure state for $n$ spins is that all but
one of the $2^n$ populations are equal, and that no coherences are
present. The density matrix then consists of an identity component and
a pure state component. The identity density matrix is not observable
in NMR since only population differences are observed, and furthermore
does not transform under unitary evolutions ($U I U^\dagger =
I$). Thus the visible signal is produced solely by the one distinct
population, corresponding to a pure state.  In product operator
notation~\cite{sorensen83}, the effective pure ground state is
proportional to $\hat{I}_z + \hat{S}_z + 2\hat{I}_z\hat{S}_z$ for two
spins, to $\hat{I}_z + \hat{S}_z + \hat{R}_z + 2\hat{I}_z\hat{S}_z +
2\hat{I}_z\hat{R}_z + 2\hat{S}_z\hat{R}_z +
4\hat{I}_z\hat{S}_z\hat{R}_z$ for three spins, and so
forth. Figure~\ref{fig:effect_pure} shows how the effective pure state
preparation is manifest in the spectrum of one of 5 coupled spins.
Characteristic of effective pure (basis) states is that only one line
survives in each multiplet.

Three methods are known for preparing effective pure states starting from
thermal equilibrium.

(1) {\em Logical labeling}~\cite{gershenfeld97,vandersypen99} consists
    of applying a pulse sequence which rearranges the thermal
    populations such that a subset of the spins is in an effective
    pure state, conditioned upon the state of the remaining spins.
    Then the computation is carried out within this embedded
    subsystem~\cite{suter86}. For example, the Boltzman populations
    for the states $\{|000\>,$ $|001\>,$ $|010\>,$ $|011\>,$ $|100\>,$
    $|101\>,$ $|110\>,$ $|111\>\}$ for a homonuclear three-spin system
    deviate from the uniform background by $\{3a,a,a,-a,a,-a,-a,-3a\}$
    respectively, where $a= \frac{1}{2^3} \frac{\hbar \omega}{2 k_B T}
    \ll 1$. After rearranging the populations for the eight spin
    states as $\{3a,-a,-a,-a,a,a,a,-3a\}$, the last two qubits are in
    an effective pure state conditioned upon the first qubit being
    $|0\>$.  As the total number of qubits $n$ in the molecule
    increases, the relative fraction of effective pure qubits goes to
    $1$, but the preparation sequence becomes complex quite rapidly
    for large $n$ and the signal strength scales as $n/2^n$.

(2) {\em Temporal averaging}~\cite{knill98} is similar to phase
    cycling ({\em see \bf Phase Cycling}), since it consists of
    adding up the spectra of multiple experiments. However, instead of
    changing just the phase of some pulses, each experiment starts off
    with a different state preparation sequence which permutes the
    populations. For two heteronuclear spins, adding together three
    experiments which yield respective population deviations
    $\{a,b,-b,-a\}$, $\{a,-b,-a,b\}$ and $\{a,-a,b,-b\}$ is equivalent
    to performing an experiment with population deviations
    $\{3a,-a,-a,-a\}$.  For arbitary $n$, at least $(2^n-1)/n$
    experiments are needed~\cite{vandersypen00b}, since the effective
    pure state is made up of $2^n-1$ product operator terms and the
    starting state, thermal equilibrium, contains $n$ terms.

(3) {\em Spatial averaging}~\cite{cory98a} uses a pulse sequence
    containing magnetic field gradients ({\em see \bf Field Gradients
    \& Their Application}) to equalize all the populations but the
    ground state population. Only one experiment is involved, but the
    preparation sequence quickly becomes unwieldy for large spin
    systems and the signal strength decreases exponentially with $n$.

To date, temporal and spatial averaging have been the most popular
choices for preparing effective pure states. Several hybrid
schemes~\cite{knill98,knill00} have also been developed which trade
off complexity of the preparation steps for the number of experiments.
Nonetheless, all these state preparation schemes have in common that\\

{\em creating effective pure states incurs an exponential cost either
in the signal strength or in the number of experiments involved.}\\

Such an exponential overhead is of course not tolerable for quantum
computations. The reason for this cost is that effective state
preparation techniques simply select out the signal from the ground
state population present in thermal equilibrium, and the fraction of
the molecules in the ground state is proportional to $n/2^n$.
computation.

A significant breakthrough by Schulman and Vazirani resulted in a
method to cool a subset of the spins in a molecule down to the ground
state without any exponential overhead~\cite{schulman98,chang00}.  The
idea is to redistribute the entropy of the spins so that some have
zero entropy (pure state) while the entropy of the remaining spins
increases.  In a sense, this method is an advanced polarization
transfer scheme. Surprisingly, both the length of the cooling pulse
sequence and the number of spins that must be sacrificed increase only
about linearly with the number of pure spins desired.  Furthermore,
the cooling algorithm approaches the entropic bound in the limit of
large numbers of spins. However, with initial polarization $\alpha \ll
1$, a molecule with at least $k/\alpha^2$ spins is needed to obtain
$k$ pure spins. This method is therefore impractical when starting
from equilibrium at room temperature, with $\alpha \approx
10^{-5}$. Nevertheless, a combination of hyperpolarization techniques
and the Schulman-Vazirani scheme may some day become practical.  In
any case, despite the exponential cost incurred when preparing
effective pure states,\\

{\em the highly random initial state represents no fundamental
obstacle to scalable quantum computation.}

\subsection{Read-out}

Traditionally in NMR spectroscopy, only one nucleus is observed. In
QC, however the concept of a single observe channel and one or more
decoupler channels does not apply:\\

{\em the output of a quantum computation is the final state of one or
several spins. The final states of each of the output spins must thus
be read out.}\\

If each of the output spins ends up in $|0\>$ or $|1\>$ (or in reality
in the effective pure state corresponding to $|0\>$ or $|1\>$), the
answer can be read-out directly by applying a pulse which rotates the
spin from $\pm z$ to $\pm x$. With properly referenced receiver phase
settings, the spectrum for each output spin then consists of either
absorption or emission lines, indicating whether the output value of
the corresponding bit is ``0'' or ``1'' (Fig.~\ref{fig:output}).

If the output state is a superposition state, the situation is a bit
more complicated. For a single (as opposed to an ensemble) quantum
computer subject to a ``hard'' measurement (assumed in
Section~\ref{sec:qubits}), the superposition ``collapses'' to one of
the terms in the superposition, with probabilities given by the
square of the amplitude of each term.  In contrast, \\

{\em measurements in NMR give a (bit-wise) ensemble averaged read-out.}\\

The output state of eq.~\ref{eq:output_qft} serves as an example: half
of the molecules in the ensemble collapse to $|0\>$ ($|000\>$), while
the other half collapses to $|4\>$ ($|100\>$). In other words, spins
$2$ and $3$ always end up in ``0'' so their spectral lines are
absorptive; in contrast, the signal of spin $1$ averages to zero
because there are equally many molecules in which spin $1$ ends up in
``0'' as in ``1''. It is not clear that such bit-wise averages of
probabilistic output states are generally sufficient to solve the
problem of interest. For Shor's period-finding algorithm, this problem
can be circumvented~\cite{gershenfeld97} by performing the classical
post-processing steps (Section~\ref{sec:algorithms}) on the quantum
computer using some ancillae qubits --- any classical computation can
also be done on a quantum computer~\cite{chuang00}. In this way, the
output state becomes the period $r$ for all the molecules in the
ensemble (as opposed to an average over all the multiples of $N/r$)
and the measurement result becomes deterministic.

Instead of recording the signal of each of the output spins, it is
sometimes possible to use the extra information provided by the line
splittings due to $J$ couplings to derive the output state of several
qubits from the spectrum of a single spin. Since each of the lines in
the multiplet can be identified with specific states of the other
spins (as in Fig.~\ref{fig:effect_pure}), the presence or absence of
each line in the multiplet gives information about the state of the
other spins (Fig.~\ref{fig:output}).

Finally, while the spectra of a few select spins suffice to obtain the
answer to a computation, the full density matrix conveys much more
information.  This extra information can be used to expose the
presence of errors such as multiple quantum coherences not visible in
the single output spectra and furthermore is a useful tool for
debugging pulse sequences.  The procedure for reconstructing the
density matrix is called quantum state
tomography~\cite{chuang98a,chuang98b,chuang98c}. It consists of
repeating the computation multiple times, each time looking at the final
state of the spins after applying different sets of read-out pulses which
rotate different elements of the density matrix to observable
positions. However, since this procedure involves on the order of
$4^n$ experiments, it is practical only for experiments involving a
few spins.

\subsection{State-of-the-art and outlook}

To date, only very simple demonstrations of quantum algorithms, all
using liquid-state NMR techniques, have been carried out. Variations
of Grover's algorithm have been demonstrated with
two~\cite{chuang98b,jones98b,jones99b} and three
qubits~\cite{vandersypen00a}, the Deutsch-Jozsa algorithm with
two~\cite{chuang98a,jones98a}, three~\cite{linden98} and
five~\cite{marx00} qubits, and the period-finding algorithm with five
qubits~\cite{vandersypen00b}. A quantum simulation has been
implemented with two~\cite{somaroo99} and three~\cite{tseng00} qubits,
and quantum error detection and correction with two~\cite{leung99} and
three~\cite{cory98b} qubits. Finally, a 7-spin coherence has been
created and observed~\cite{knill00} (see Ref. ~\incite{jones00a} for
additional references).  While these experiments indeed demonstrate the
principles of quantum information processing, they all involve far
fewer qubits than would be needed to solve a problem beyond the reach
of classical machines.

Despite the rapid progress in recent years, scaling liquid state NMRQC
to tens or hundreds of qubits may be impractical for several reasons,
although none of them appear fundamental. In particular, as the number
of qubits increases: (1) the strength of the signal selected with
current state initialization techniques decreases
exponentially~\cite{gershenfeld97,cory97,warren97}, but
Schulman-Vazirani cooling~\cite{schulman98} does not suffer any such
exponential overhead (Section~\ref{sec:implement}); (2) the chemical
shift separations unavoidably become smaller, but Lloyd~\cite{lloyd93}
showed that for universal quantum computation it suffices to have a
linear chain $d-abc-abc-\ldots-abc$, in which only nearest neighbour
spins are coupled and with only four distinct chemical shifts
$\delta_a$, $\delta_b$, $\delta_c$ and $\delta_d$ ; (3) $J$-couplings
become smaller or even unresolved, but this can be circumvented by
swapping spin states (Section~\ref{sec:implement})~\cite{lloyd93}.
While these obstacles are not fundamental, the solutions all
make the pulse sequences much longer. This would require increasingly
longer $T_2$'s and $T_1$'s for larger molecules, while in practice the
$T_2$'s and $T_1$'s tend to become shorter ({\em see \bf Relaxation:
An Introduction}).

NMRQC has also brought up new theoretical issues.
(1) Since only ensemble averaged results are available because of the
    large number of molecules in a sample tube, some information is
    lost that would be available in an idealized quantum computer such
    as a single molecule at zero degrees Kelvin. For the known quantum
    algorithms, this information can be retrieved by performing
    classical post-processing steps on the quantum computer
    (Section~\ref{sec:algorithms}).
(2) Since the density matrix of nuclear spins at room temperature is
    very close to the identity matrix, it is not possible to produce
    genuinely entangled states between the nuclear spins in small
    thermally polarized molecules in liquid
    solution~\cite{braunstein99}. This observation has sparked a
    stimulating debate about the ``quantumness'' of NMR, because it
    implies that each of the states produced in NMRQC experiments so
    far is classical. However, all attempts to describe the dynamics
    of a set of coupled spins by an efficient classical model have
    been unsuccesful. It is thus conjectured that even though the
    states are classical, the dynamics of the spins is truly quantum
    mechanical~\cite{schack99}, a proposition which will appear
    obvious to most NMR spectroscopists. In fact, it has also been the
    starting point of this introduction to NMR quantum computing.

\section{Summary and conclusions}

In many respects, liquid state NMR provides an ideal test bed for
elementary quantum computations. The degree of control over the
evolution of multiple coupled qubits --- the result of 50 years of
technology development --- the long relaxation times of nuclear spins
and a set of new insights~\cite{gershenfeld97,cory97} made it possible
to perform certain computations in {\em fewer steps} than is possible
using any classical machine. This is in itself a remarkable
achievement.

It is unlikely that liquid state NMR could ever be used to solve
problems {\em faster} than any classical machine, but it has already
inspired many other NMR based proposals for quantum computing.
Liquid-crystal solvents have been used to partly reintroduce
dipole-dipole couplings ({\em see \bf Liquid Crystals: General
Considerations}) to speed up the gate time and increase the number of
gates possible within the coherence time~\cite{yannoni99}.  New
molecular architectures based on liquid crystal solutions are now
being investigated.  Solid-state NMR near zero Kelvin
could circumvent the state initialization problem, but also poses new
questions in terms of addressability and coherence
times ({\em see \bf Internal Spin Interactions \& Rotations in
Solids}). Several approaches to solve these issues have been
proposed~\cite{cory00,yamaguchi99}.  Another proposal which received
much attention consists of doing NMR on impurity atoms placed in a
linear or two-dimensional array, with chemical shifts and couplings
controlled by electrodes placed on top of and in between the impurity
atoms~\cite{kane98}.

Furthermore, there is a plethora of very different experimental
approaches to building quantum computers (for an extensive review, see
Ref. ~\incite{fortschritte00}).  Four trapped ions have recently been
entangled~\cite{sackett00} and a single quantum logic gate has been
implemented between two photons coupled with each other via
interactions with an atom in a cavity.  In the long run, the most
scalable approaches may be those based on solid-state technology, such
as electron spins in quantum dots or magnetic fluxes in
SQUIDs~\cite{fortschritte00}.

It is clear that none of these proposals will be easy to implement
since they all require substantial and innovative development of
technology.  The success of any approach will depend on the ratio of
the coherence time to the gate duration, i.e. how many gates can be
completed within the coherence time, and on the achievable degree of
quantum control.  

Many of the problems in quantum control are similar for different
experimental systems, and we therefore expect that other quantum
computer implementations will benefit from the ideas, concepts and
solutions which arise from liquid state NMR experiments.  In addition,
we hope that some of the techniques developed within the context of
quantum computation may find more general application in NMR.

The possible payoff for successful quantum computing is tremendous: to
solve problems beyond the reach of any classical computer. It is not
clear at this point whether quantum computers will fulfill this
promise, but in any case quantum computing has already provided an
exciting new perspective on NMR and, more broadly, on the
connection between physics, information and computation.

\begin{figure}
\vspace*{2cm}
\caption{The evolution of one of two coupled heteronuclear spins
during an INEPT type pulse sequence, when the other spin is up (solid
line) or down (dashed line). The rotating frame is set on resonance
with the first spin so there is no need to refocus chemical shift. The
usual read-out pulse is left out.  The same pulse sequence can be
applied to two homonuclear spins using spin-selective pulses.}
\label{fig:inept}
\end{figure}
\vspace*{2cm}

\begin{figure}
\caption{(a) Input and output states for the INEPT pulse sequence and
(b) for the corresponding {\sc cnot} gate.}
\label{tab:inept}
\end{figure}
\vspace*{2cm}

\begin{figure}
\caption{Schematic diagram of the main steps in quantum algorithms for period-finding.}
\label{fig:shor_structure} 
\end{figure}
\vspace*{2cm}

\begin{figure}[h]
\caption{(a) Spectrum of pentafluorobutadienyl
cyclopentadienyldicarbonyliron complex in thermal equilibrium. (b) The
same spectrum after preparing an effective pure state $\ket{00000}$.
(Reproduced from Ref. ~\protect\incite{vandersypen00b} by permission of the
American Physical Society)}
\label{fig:effect_pure}
\end{figure}
\vspace*{2cm}

\begin{figure}
\caption{Output spectra of the proton (left) and carbon (right) spins
of $^{13}$CHCl$_3$ (disolved in a liquid crystal solution) for four
different executions of Grover's search algorithm. Only the real part
of the spectra is shown, and frequencies are relative to $\nu_H$ and
$\nu_C$. A positive or negative line in the spectrum indicates that
the corresponding spin was in $\ket{0}$ or $\ket{1}$ before the
read-out pulse. Furthermore, the position of the line in the spectrum
of one spin also reveals the state of the other spin. For example, if
the $^1$H line is at $\nu_H - J_{CH}/2$, the $^{13}$C spin is in
$\ket{0}$; a $^1$H line at $\nu_H + J_{CH}/2$ indicates the $^{13}$C
spin is in $\ket{1}$. Thus, the state of the two qubits for each of
the four cases is (from top to bottom) $\ket{00}$, $\ket{01}$,
$\ket{10}$ and $\ket{11}$.  (Reproduced from
Ref.~\protect\incite{yannoni99} by permission of the American
Institute of Physics)}
\label{fig:output} 
\end{figure}

\clearpage

\begin{center}

\noindent \mbox{\psfig{file=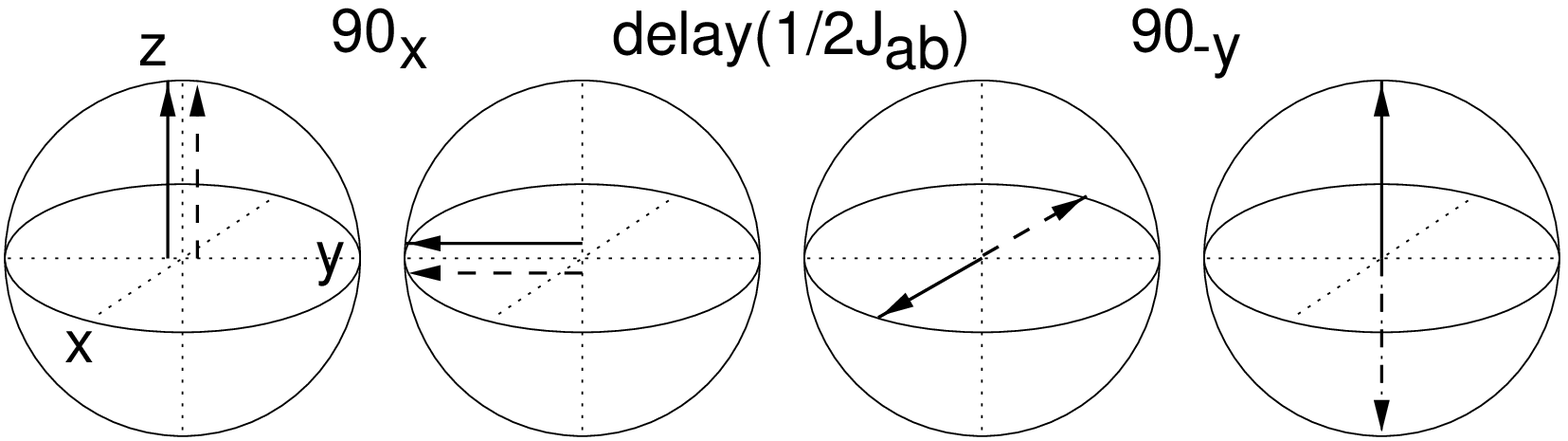,width=3.4in}} \\
Figure 1\\
\vspace*{2cm}

\begin{tabular}{cc|cc}
\multicolumn{2}{c|}{in} & \multicolumn{2}{c}{\hspace*{0.ex} out} \\ \hline
\ua & \ua & \ua & \ua \\
\ua & \da & \ua & \da \\
\da & \ua & \da & \da \\
\da & \da & \da & \ua
\end{tabular}
\hspace*{0.2cm}
\begin{tabular}{cc|cc}
\multicolumn{2}{c|}{in} & \multicolumn{2}{c}{\hspace*{0.ex} out} \\ \hline
0 & 0 & 0 & 0 \\
0 & 1 & 0 & 1 \\
1 & 0 & 1 & 1 \\
1 & 1 & 1 & 0 \\
\end{tabular}\\
\hspace*{0ex} (a) \hspace*{6.5ex} (b) \\
Figure 2\\
\vspace*{2cm}

\noindent \mbox{\psfig{file=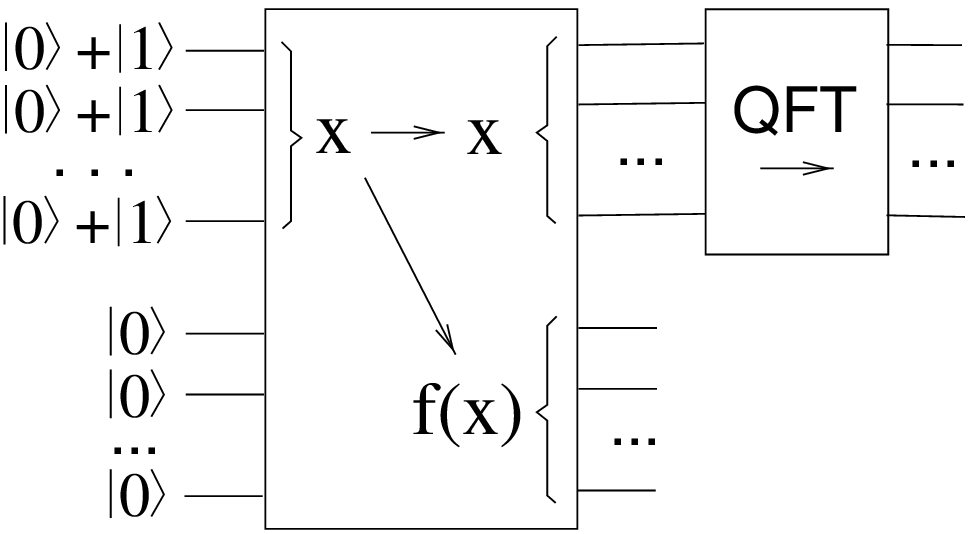,width=2in}}\\
Figure 3\\
\vspace*{2cm}

\clearpage

\noindent \mbox{\psfig{file=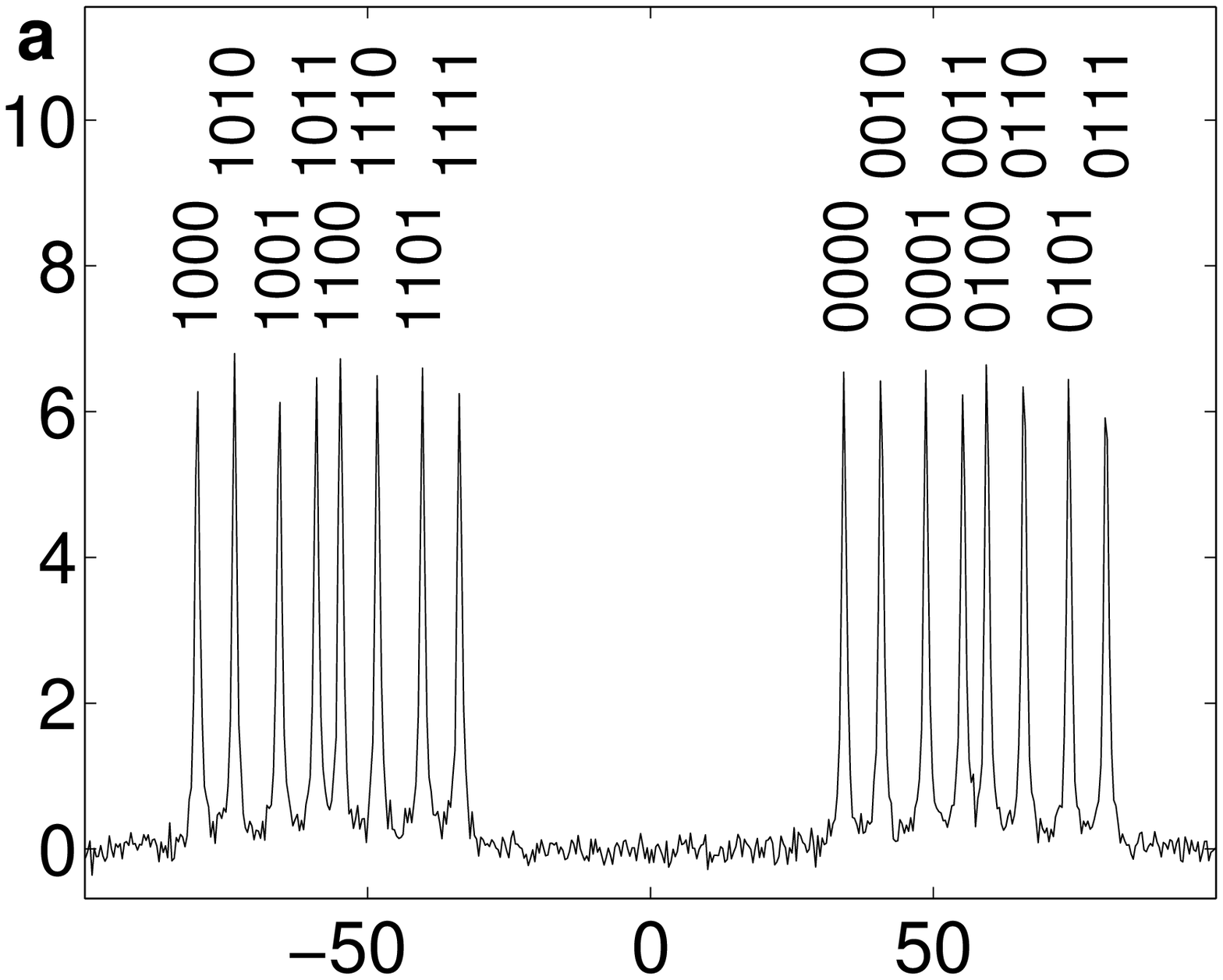,height=1.8in}}
\mbox{\psfig{file=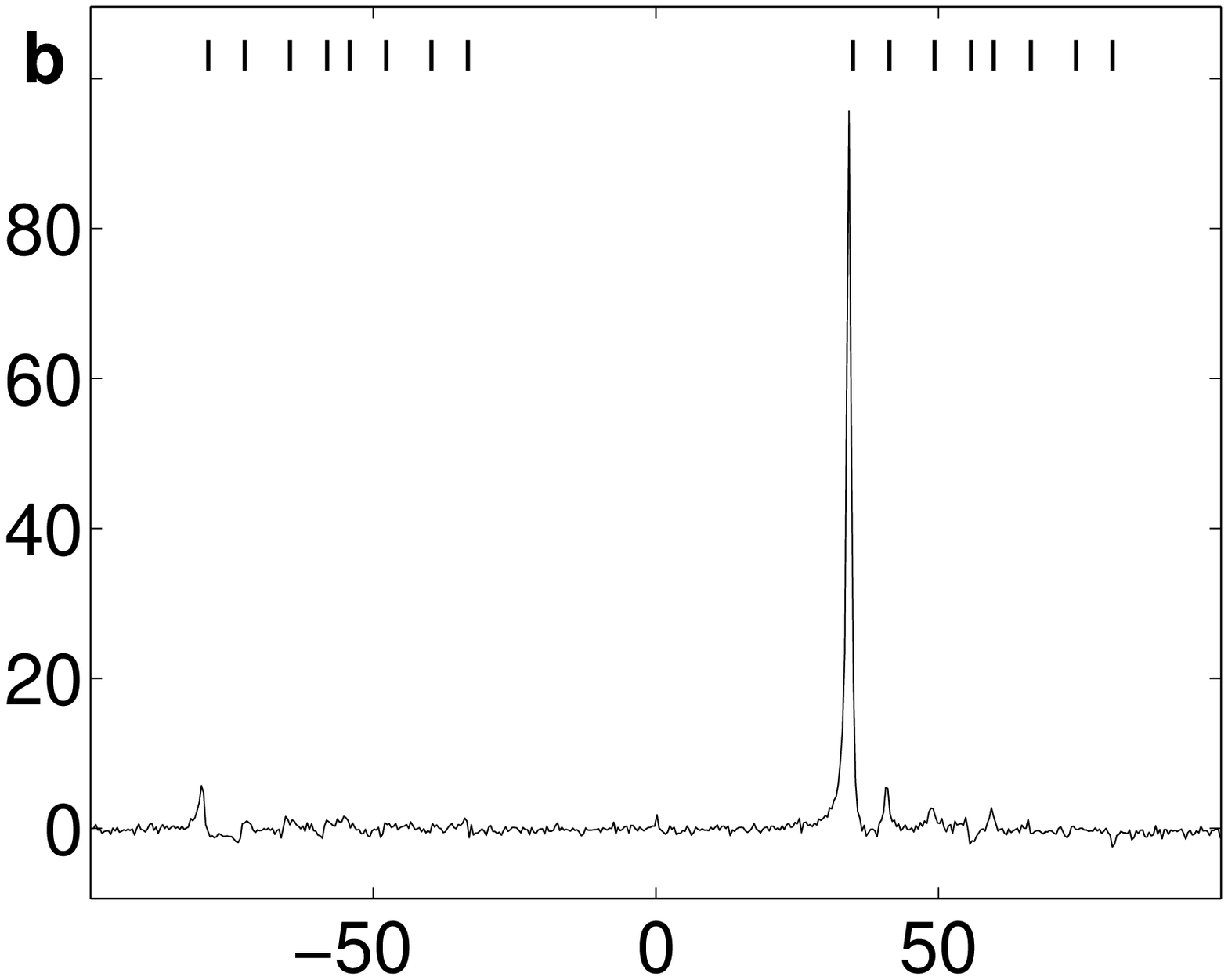,height=1.8in}}\\
Figure 4\\
\vspace*{2cm}

\noindent 
$$
\begin{array}{cc}
\mbox{\psfig{file=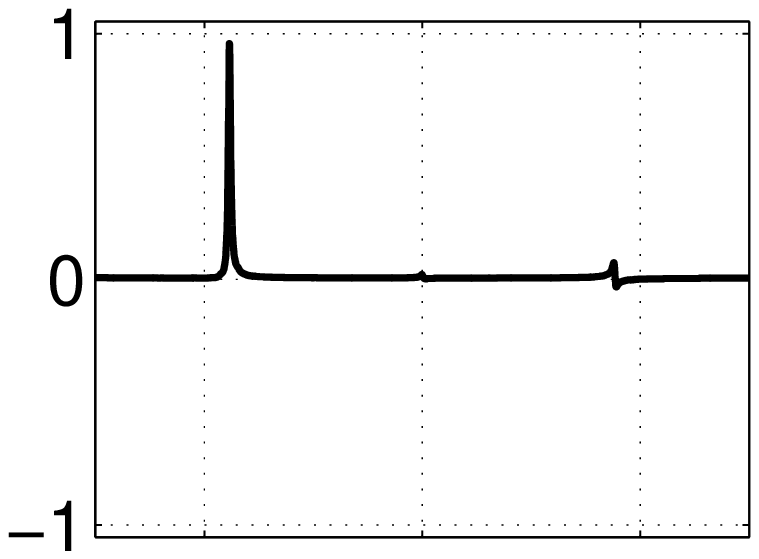,width=1.4in}} &
\mbox{\psfig{file=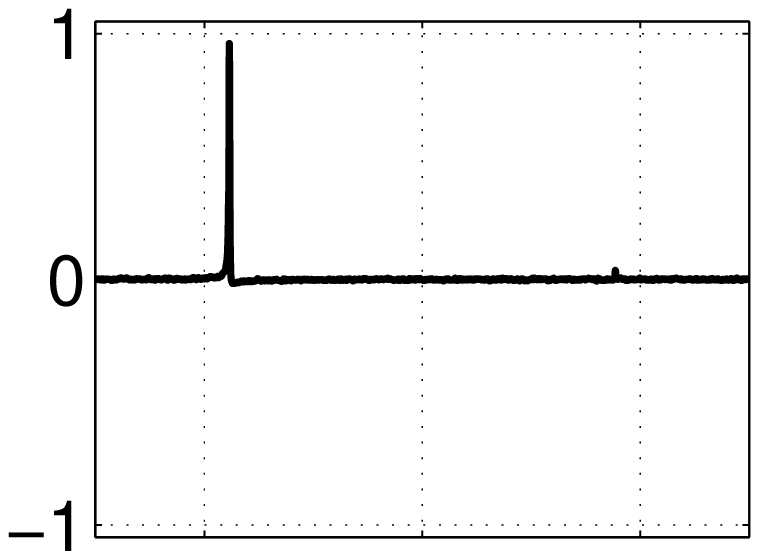,width=1.4in}} \\
\mbox{\psfig{file=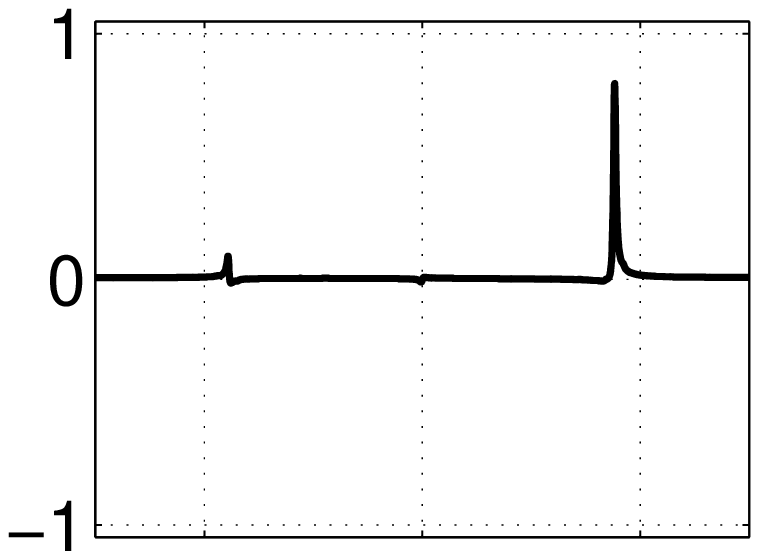,width=1.4in}} &
\mbox{\psfig{file=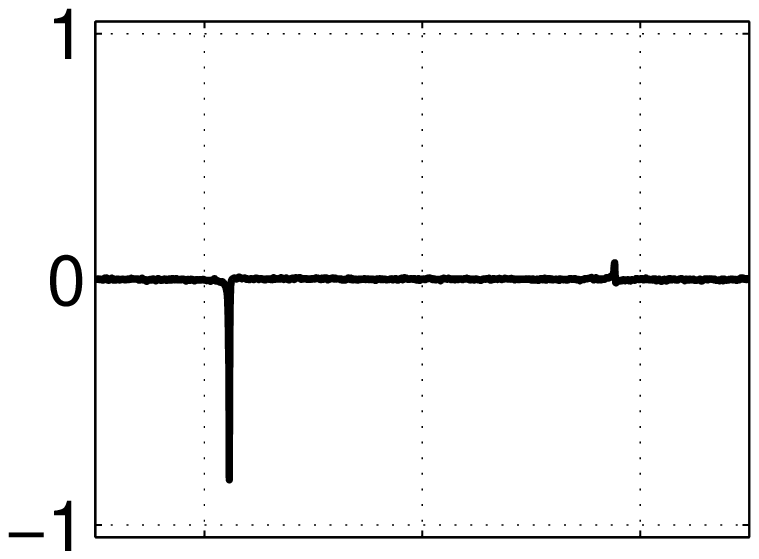,width=1.4in}} \\
\mbox{\psfig{file=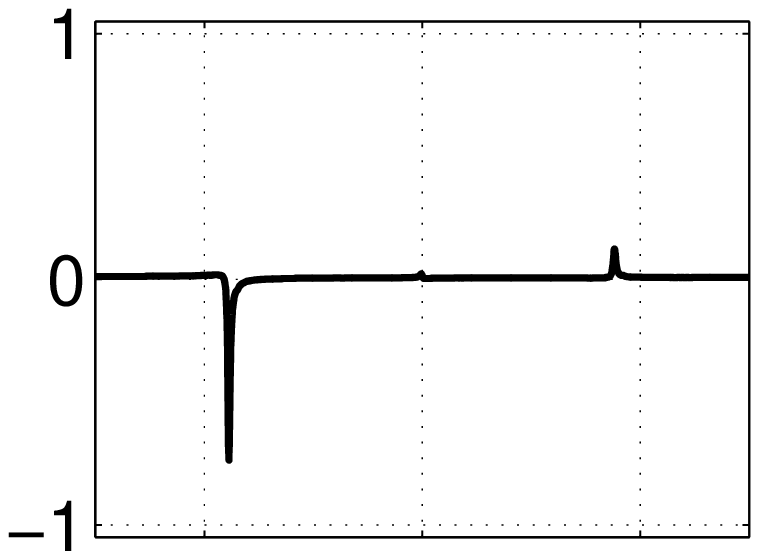,width=1.4in}} & 
\mbox{\psfig{file=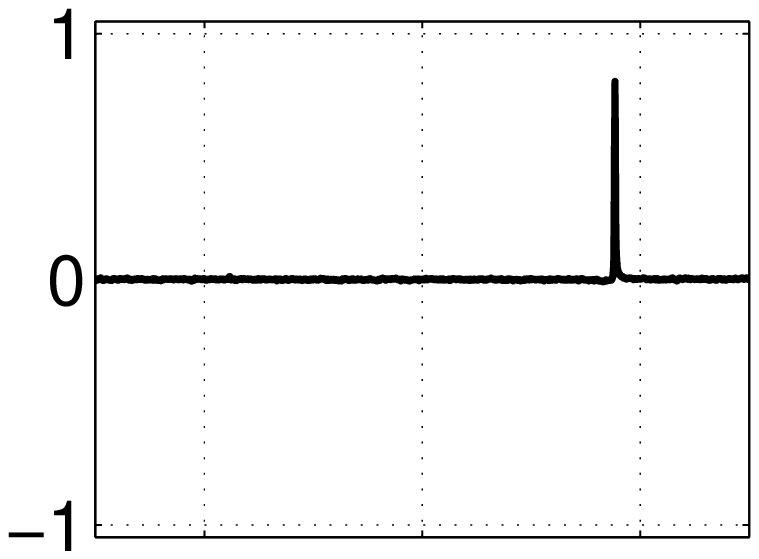,width=1.4in}} \\
\mbox{\psfig{file=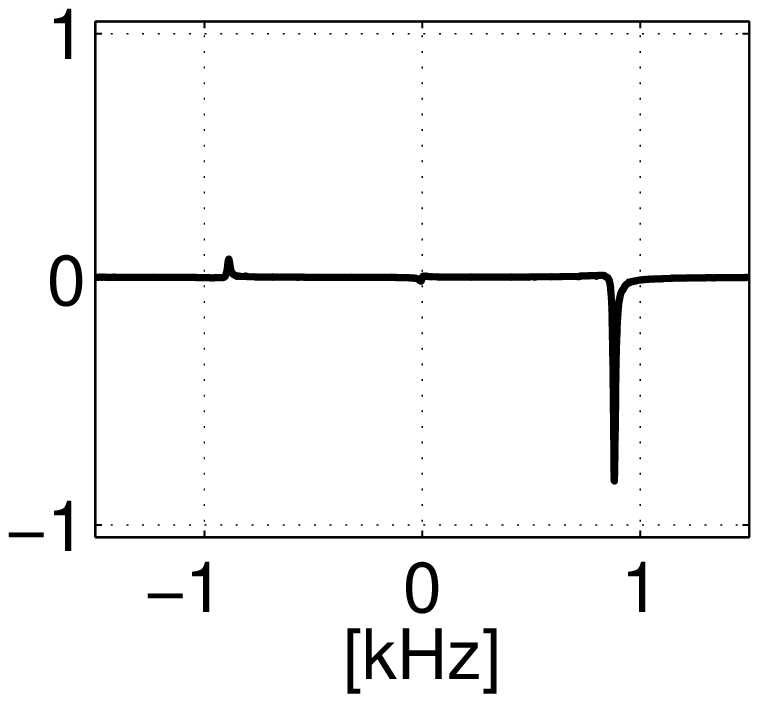,width=1.4in}} & 
\mbox{\psfig{file=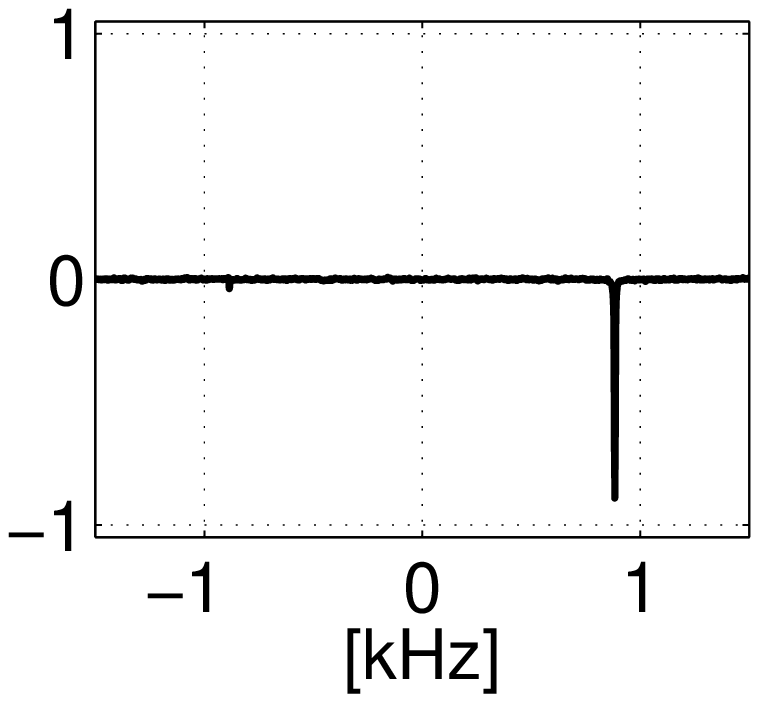,width=1.4in}} \\
\end{array}
$$
Figure 5\\

\end{center}

\end{document}